\documentclass[structabstract]{aa}  
\usepackage{graphicx}
\usepackage{url}
\usepackage{txfonts}
\begin{document}

   \title{Detection of a multi-shell planetary nebula around the hot subdwarf O-type star 
2MASS\,J19310888+4324577}

   \author{A. Aller\inst{1}, L.~F. Miranda\inst{1,2}, A. Ulla\inst{1},
     R.~V\'azquez\inst{3}, P.~F. Guill\'en\inst{3}, L. Olgu\'{\i}n\inst{4}, 
C. Rodr\'{\i}guez-L\'opez\inst{1,5}, P. Thejll\inst{6}, R. Oreiro\inst{5}, 
     M. Manteiga\inst{7} \and E. P\'erez\inst{1}}

   \institute{Departamento de F\'isica Aplicada, Universidade de Vigo,
              Campus Lagoas-Marcosende s/n, E-36310 Vigo, Spain \\
              \email{alba.aller@uvigo.es, ulla@uvigo.es, estherperez@edu.xunta.es}
         \and
             Consejo Superior de Investigaciones Cient\'ificas, Serrano 117,
             E-28006 Madrid, Spain \\
              \email{lfm@iaa.es}
         \and 
             Instituto de Astronom\'{\i}a, Universidad Nacional Aut\'onoma
             de M\'exico, Apdo. Postal 877, 22800 Ensenada, B.C., Mexico \\ 
              \email{vazquez@astro.unam.mx, fguillen@astro.unam.mx}
          \and   
             Departamento de Investigaci\'on en F\'{\i}sica, Universidad de Sonora,
             Blvd. Rosales Esq. L.D. Colosio, Edif. 3H, 83190 Hermosillo,
             Son. Mexico \\
              \email{lorenzo@astro.uson.mx} 
           \and   
             Instituto de Astrof\'{\i}sica de Andaluc\'{\i}a - CSIC, Glorieta de
             la Astronom\'{i}a, s/n. E-18008, Granada, Spain \\ 
              \email{roreiro@iaa.es, crl@iaa.es} 
          \and   
             Danish Climate Centre at the Danish Meteorological Institute,
             Lyngbyvej 100, DK-2100 Copenhagen, Denmark \\
              \email{pth@dmi.dk}
          \and   
          Departamento de Ciencias de la Navegaci\'on y de la Tierra,
          Universidade da Coru\~na, Rua de Maestranza, E-15001 A Coru\~na,
          Spain \\
              \email{manteiga@udc.es}
             }

   \date{Received; accepted }

 
\abstract
   {The origin of hot subdwarf O-type stars (sdOs) remains unclear since their
     discovery in 1947. Among others, a post-Asymptotic Giant Branch
     (post-AGB) origin is possible for a fraction of sdOs.}
   {We are involved in a comprehensive ongoing study to search for and to
     analyze planetary nebulae (PNe) around sdOs with the aim of establishing the
     fraction and properties of sdOs with a post-AGB origin.}
   {We use deep H$\alpha$ and [O\,{\sc iii}] images of sdOs to detect nebular
     emission and intermediate resolution, long-slit optical spectroscopy of the 
     detected nebulae and their sdO central stars. These data are complemented with 
other observations (archive images, high-resolution, long-slit spectroscopy) for further 
analysis of the detected nebulae.}
   {We report the detection of an extremely faint, complex PN around
     2MASS\,J19310888+4324577 (2M1931+4324), a star classified as sdO in a binary system. 
     The PN shows a bipolar and an elliptical shell, whose major axes are oriented 
     perpendicular to each other, and high-excitation structures outside the two
     shells. WISE archive images show faint, extended emission at 12\,$\mu$m
     and 22\,$\mu$m in the inner nebular regions. The internal nebular
     kinematics, derived from high resolution, long-slit
spectra, is consistent with a bipolar and a 
cylindrical/ellipsoidal shell, in both cases with the main axis mainly perpendicular
to the line of sight. The nebular spectrum only exhibits H$\alpha$, H$\beta$ and 
     [O\,{\sc iii}]$\lambda$$\lambda$4959,5007 emission lines, but suggests a very low-excitation ([O\,{\sc
       iii}]/H$\beta$ $\simeq$1.5), in strong contrast with the absence of
     low-excitation emission lines. The spectrum of
     2M1931+4324 presents narrow, ionized helium absorptions that confirm the
     previous sdO classification and suggest an effective temperature
     $\geq$60000\,K. The binary nature of 2M1931+4324, its association with a complex PN, and
     several properties of the system provide strong support for the idea that binary central stars
     are a crucial ingredient in the formation of complex PNe.}
   {}


   \authorrunning{Aller et al.} 
   \titlerunning{Detection of a planetary nebula around the sdO 2M1931+4324}
   \maketitle
%


\section{Introduction}

Hot subdwarf O-type stars (sdOs) are low--mass, blue subluminous objects
evolving towards the white dwarf phase. In the HR diagram they are 
located in a region with effective temperature ({\it T}$_{\rm eff}$) between
$\simeq$40000\,K and $\simeq$100000\,K,  
and surface gravity (log({\it g})) between $\simeq$4.0 and $\simeq$6.5. Since their discovery by 
Humason \& Zwicky (1947), the origin of sdOs has been difficult to 
establish because this region of the HR diagram is crossed by evolutionary tracks of
post-Asymptotic Giant Branch (post-AGB), post-Red Giant Branch 
and post-Extended Horizontal Branch stars. In addition, binary stars scenarios are 
also possible (e.g., Napiwotzki 2008; Heber \cite{Heber_2009}).

     \begin{figure*}
   \centering
  \includegraphics[width=\textwidth]{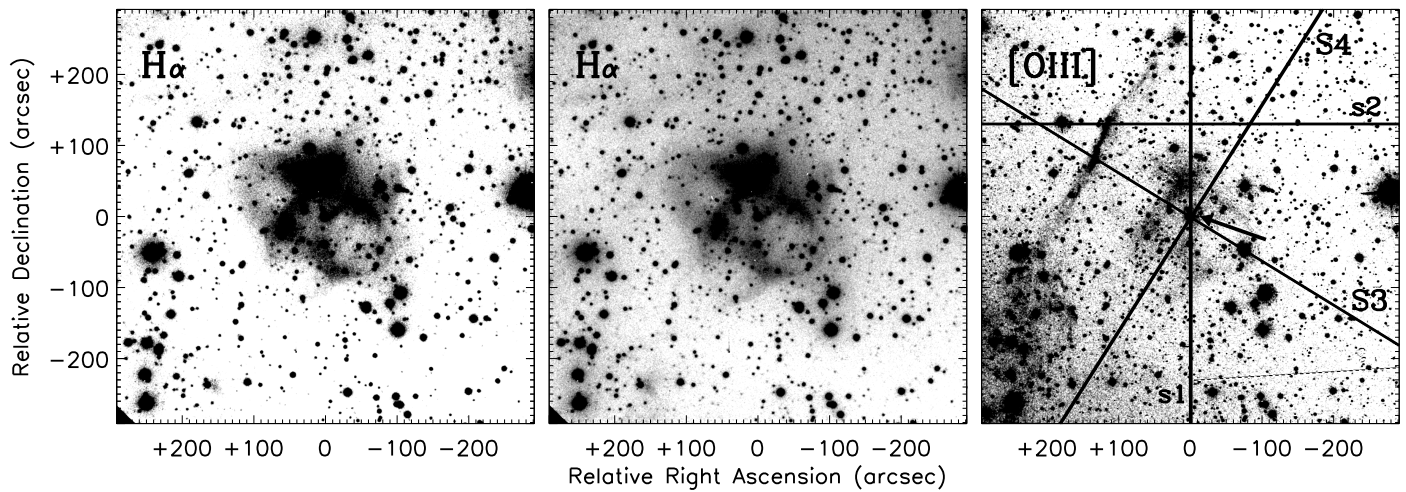}
      \caption{Grey-scale reproductions of the H$\alpha$ (left and center) and
        [O\,{\sc iii}] (right) images of 2M1931+4324. Grey levels are
        linear and two different levels are used to show the H$\alpha$
        image. The slits s1 and s2 (S3 and S4) used for intermediate (high) resolution, long-slit spectroscopy 
        are plotted on the [O\,{\sc iii}] image (slit width not to scale) in which 2M1931+4324 is also arrowed.}
   \end{figure*}

     \begin{figure*}
   \centering
  \includegraphics[width=\textwidth]{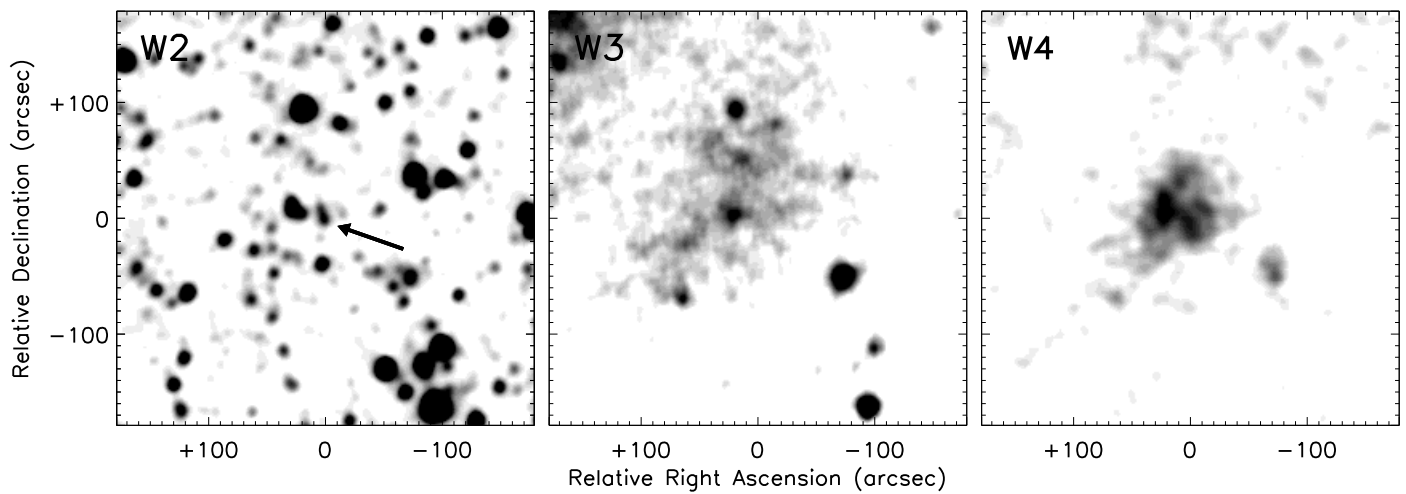}
  \caption{Grey-scale reproductions of the WISE\,2 (4.6\,$\mu$m), WISE\,3 (12\,$\mu$m), and WISE\,4 (22\,$\mu$m)
    images of 2M1931+4324. The grey levels are linear and arbitrary, and 
    have been chosen to emphasize the more relevant nebular features. 2M1931+4324 is
    arrowed in the WISE\,2 image while it is not detected in the WISE\,3 and
    WISE\,4 images.}
   \end{figure*}

The association of sdOs with planetary nebulae (PNe) is essential to obtain the 
fraction of sdOs that have a post-AGB origin. As of now, only a few sdO+PN
associations have been reported, including the
``classical'' sdOs LSE\,125 and LSS\,2018  (Drilling \cite{Drilling_1983}),
LSS\,1362 (Heber \& Drilling \cite{Heber_Drilling_1984}), and RWT\,152
(Pritchet \cite{Pritchet_1984}), as well as several PNe whose central stars have been classified as 
sdOs (e.g., Rauch et al. 2002; Montez et al. 2010; Weidemann \& Gamen 2011). 
On the other hand, only a few sdOs have been searched for a surrounding PN. M\'endez et 
al. (\cite{Mendez_etal88}) carried out long-slit spectroscopy of 12 sdOs to search for extended 
emission. The few detected cases were identified by the authors as diffuse ionized regions in the
Galactic disk, and M\'endez et al. recommended an image survey to analyze the morphology of the
detected extended emissions. Although morphology should not be used as unique criterium to identify PNe, 
it is true that most PNe present axisymmetric shells that are not usually observed in other objects 
(see, e.g., Miranda et al. 2009). An imaging survey was carried out by Kwitter et
al. (\cite{Kwitter_etal89}) who obtained narrow-band images of 14 sdOs (and long-slit spectra 
of 42 sdOs) although only with a positive and already known detection (RWT\,152). We note that the
instrumental characteristics of this imaging survey (relatively short exposure
times, small field of view) may have prevented from detecting large and/or
very faint PNe. In fact, if sdO+PN associations are common, 
these PNe should be very faint as only a few can be recognized by
a simple visual inspection in the Palomar Observatory Sky Surveys (POSS). Therefore,
deeper imaging surveys are required to prove the sdO+PN association.

     \begin{figure*}
   \centering
  \includegraphics[width=\textwidth]{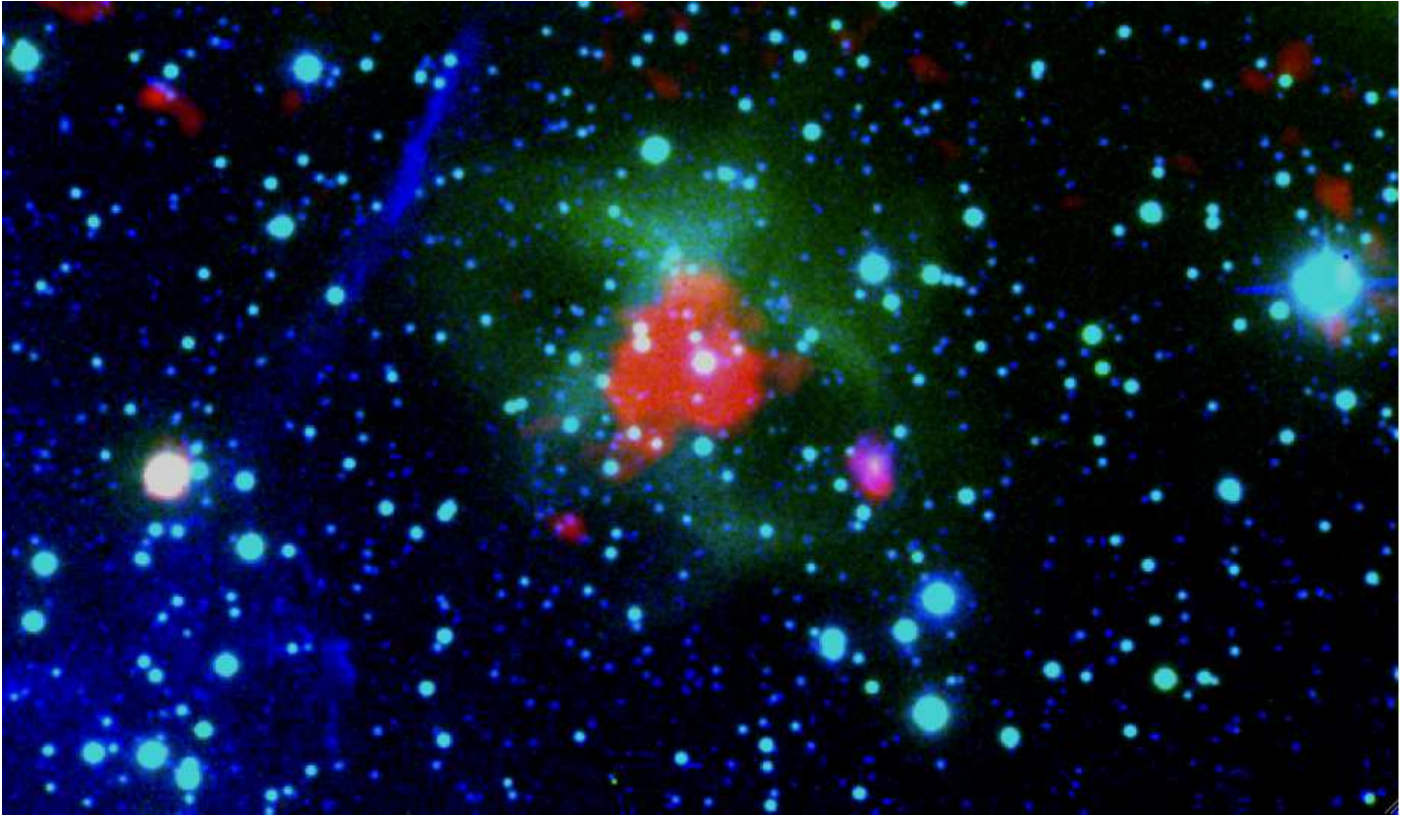}
      \caption{Colour-composite image of PN\,G\,075.9+11.6. Red corresponds to
        WISE\,4 (22\,$\mu$m), green to H$\alpha$, and blue to [O\,{\sc iii}]. North is up, east to 
the left. The size of the field shown is 10$\farcm$5 x 6\farcm1. Note that the
red knot near the tip of the southwestern 
bipolar lobe is probably a galaxy that is also observed in the optical images.}
   \end{figure*}

We are involved in a program to search for and to analyze sdO+PN
associations by means of deep direct images and long-slit spectroscopy. In this
framework, we present 2MASS\,J19310888+4324577 (hereafter 2M1931+4324; 
$\alpha$(2000.0) = 19$^h$\,31$^m$\,8$\fs$9, $\delta$(2000.0) = +43$^{\circ}$\,24$'$\,58$''$; 
{\it l} = 075$\fdg$99, {\it b} = 11$\fdg$6) 
which has been spectroscopically classified as an sdO by $\O$stensen et
al. (\cite{Ostensen_etal2010}). Recently, Jacoby (2011, private communication) 
and Jacoby et al. (2012) have analyzed photometric {\it Kepler} data
of 2M1931+4324 (Kepler\,ID\,7755741\footnote{\url{http://archive.stsci.edu/kepler/data_search/search.php}}) 
and found that it is a regular variable with a period of
$\lesssim$ 3 days, suggesting that the star is irradiating a companion. A simple visual inspection of 
the POSS red and blue plates does not reveal extended emission around 2M1931+4324. However, 
when the plates are inspected at very low intensity levels, an extremely faint nebulosity around 2M1931+4324 
can be hinted. Because of this and the sdO classification, we included 2M1931+4324 in our target list.

In this paper, we present narrow-band optical images that reveal 
for the first time the existence of a very faint multi-shell PN around
2M1931+4324. We also show mid-IR images from the WISE
archive\footnote{\url{http://irsa.ipac.caltech.edu}}, in which the nebula 
is detected. Complementary intermediate- and high-resolution, long-slit spectra 
allow us to confirm the sdO/central star nature of 2M1931+4324, to describe
the nebular emission spectrum and to analyze the internal kinematics of the nebula.


\section{Observations and results}

\subsection{Optical imaging}

A narrow-band [O\,{\sc iii}] image was obtained on 2010 August 22 with the
Wide Field Camera (WFC) at the 2.5m Isaac Newton Telescope (INT) on El Roque de los Muchachos
Observatory\footnote{The Isaac Newton Telescope is operated on the island of
  La Palma by the Isaac Newton Group in the Spanish Observatorio de El Roque de 
los Muchachos of the Instituto de Astrof\'{\i}sica de Canarias.} (La Palma,
Canary Islands, Spain). The detector was a four 2k$\times$4k--CCD mosaic with a plate
scale of 0$\farcs$33\,pixel$^{-1}$ and a field of view of
34$'$$\times$34$'$. We used an [O\,{\sc iii}] filter ($\lambda_{\rm 0}$ = 5008
\AA, FWHM = 100 \AA) to obtain three images with an exposure time of 1800\,s
each. The sky was clear with an average seeing value of $\simeq$1$\farcs$3.

A narrow-band H$\alpha$ image was obtained on 2011 July 12 with the Calar Alto
Faint Object Spectrograph (CAFOS) at the 2.2m telescope on Calar Alto
Observatory\footnote{The Centro Astron\'omico Hispano Alem\'an (CAHA) at Calar
  Alto is operated jointly by the Max-Planck Institut f\"ur Astronomie and the
  Instituto de Astrof\'{i}sica de Andaluc\'{i}a (CSIC).} (Almer\'{i}a,
Spain). A SITe 2k$\times$2k--CCD was used as detector, with a plate 
scale of 0$\farcs$53\,pixel$^{-1}$ and a circular field of view of 16$'$ in
diameter. We used an H$\alpha$ filter
($\lambda_{\rm 0}$ = 6563 \AA, FWHM = 15 \AA) to obtain two images of 1800\,s
and 2400\,s. Weather conditions were fair and seeing was $\simeq$2$''$.

The images were reduced using standard procedures for direct image within the
IRAF and MIDAS packages. 

Figure\,1 shows the H$\alpha$ and [O\,{\sc iii}] images of 2M1931+4324 that
reveal the existence of a very faint and complex nebula around the star. In H$\alpha$
two structures can be recognized: a bipolar shell with a size of $\simeq$ 4$\farcm$3$\times$1$\farcm$7
and the major axis oriented at position angle (PA) $\simeq$ 55$^{\circ}$, and
an elliptical shell with a size of $\simeq$ 5$'$$\times$1$\farcm$8 and the major
axis oriented at PA $\simeq$ 145$^{\circ}$. The polar regions of the
elliptical shell are very faint and, in fact, the elliptical shell seems to be open. 
The nebula is particularly bright in the regions where both shells cross each other, while the shells 
enclose regions of lower intensity. The nebula is very weak in [O\,{\sc iii}], suggesting low-excitation 
(see Sect. 2.3.2). Additionally, the [O\,{\sc iii}] image reveals 
a long filament outside the two shells, which extends from the north towards
the southeast of the two shells, ending in a diffuse emission region. While the diffuse emission 
could also be present in H$\alpha$, the filament is 
not detected in this line, indicating very high-excitation. We note that 2M1931+4324 is not located at
the center of the shells but displaced $\simeq$ 10$''$ towards PA $\simeq$
235$^{\circ}$, approximately coinciding with the orientation of the minor
(major) axis of the bipolar (elliptical) shell. The presence of two axisymmetric shells around a 
hot sdO star (see Sect. 2.3.1) strongly suggests a PN nature for 
the nebula. Following the designation scheme for Galactic PNe by Acker et al. (1992), we
tentatively propose the name PN\,G\,075.9+11.6 for this nebula, and we shall 
refer to it hereafter.

\subsection{Mid--infrared imaging}

In order to further investigate this PN, we have inspected the NASA's
Wide-field Infrared Survey Explorer (WISE) database. 
WISE is a space telescope that is designed to map the entire sky in four
infrared bands: 3.4\,$\mu$m (W1), 4.6\,$\mu$m (W2), 12\,$\mu$m (W3), and
22\,$\mu$m (W4) with angular resolutions of 6$\farcs$1, 6$\farcs$4, 6$\farcs$5, and 12$\farcs$0,
respectively. Because we are interested in the nebular morphology, we
retrieved the W1 to W4 images in the atlas, which have a spatial scale of
1$\farcs$375\,pixel$^{-1}$. 

Figure\,2 shows the W2, W3, and W4 images around 2M1931+4324 (W1 is not shown here). 
While no nebula can be recognized in the W1 and W2 images,
the W3 and W4 ones show nebulous emission that seems to be related to PN\,G\,075.9+11.6. 
In W3, very weak extended emission is detected in a region of $\simeq$
3$\farcm$1$\times$1$\farcm$5 in size oriented at PA $\simeq$ 145$^{\circ}$. The orientation 
is similar to that of the elliptical shell described in Sect. 2.1, although 
the nebulosity observed in W3 is displaced towards the northeast with respect to the elliptical shell 
observed in H$\alpha$. In W4, extended emission is observed in a region of
$\simeq$ 50$''$ in size at the center of the nebula and presents an irregular, knotty 
morphology. The origin of the 12\,$\mu$m and 22\,$\mu$m emission associated to PN\,G\,075.9+11.6 is 
difficult to establish without a spectrum. In principle, [S\,{\sc iv}] and
[Ne\,{\sc iii}] emission lines could contribute to the emission observed in
W3, and [O\,{\sc iv}] emission line could partially be included in that observed in W4 (see Ressler et al. 
2010). However, the absence of (prominent) high-excitation emission lines in the optical spectrum 
(e.g., [Ne\,{\sc iii}], see Sect. 2.3.2) seems to suggest that the mid-infrared
emission is due to cool dust. We also note that 2M1931+4324 itself is detected
in W1 and W2, but not in W3 and W4. 

In order to analyze the morphological relationship between the emission observed at 22\,$\mu$m (W4) 
and that in the optical images, we present in Figure\,3 a colour composite image obtained
by combining H$\alpha$, [O\,{\sc iii}], and W4. This image shows that the
H$\alpha$ emission dominates in the bipolar shell and outer regions of the
elliptical shell. The [O\,{\sc iii}] emission comes mainly from the regions
where the bipolar and elliptical shells cross each other, and it is the
dominant emission in the high-excitation filament and diffuse region. The emission at 22\,$\mu$m 
traces only a part of the northwestern border of the elliptical 
shell, and it is also detected in a region surrounding the central star. 

\subsection{Intermediate-resolution optical spectroscopy}

Intermediate-resolution, long-slit spectra were acquired on 2011 July 13 with
CAFOS. Gratings B-100 and R-100 were used to cover the 3200--6200 $\AA$ and
5800--9600 $\AA$ spectral ranges, respectively, both at a dispersion of
$\simeq$ 2\,$\AA$\,pixel$^{-1}$. The slit width was 2$''$ and spectra were
obtained at two slit positions: one (denoted s1) with the slit
oriented north--south, centered on 2M1931+4324 and with an exposure time of
3600\,s for each grism, in order to cover 2M1931+4324 and the brightest
nebular regions; and another (s2) with the slit oriented east--west, centered
2$\farcm$2 northern of 2M1931+4324 and with an exposure time of 2400\,s for each grism,
in order to register the outer filament. These slits are also shown in
Fig.\,1. The spectrophotometric standard HZ\,44 was also observed for flux
calibration. The sky was clear and seeing was $\simeq$ 2$''$. The spectra were
reduced using standard procedures for long-slit spectroscopy within the IRAF
and MIDAS packages.

   \begin{figure}
   \centering
   \includegraphics[width=8.5cm]{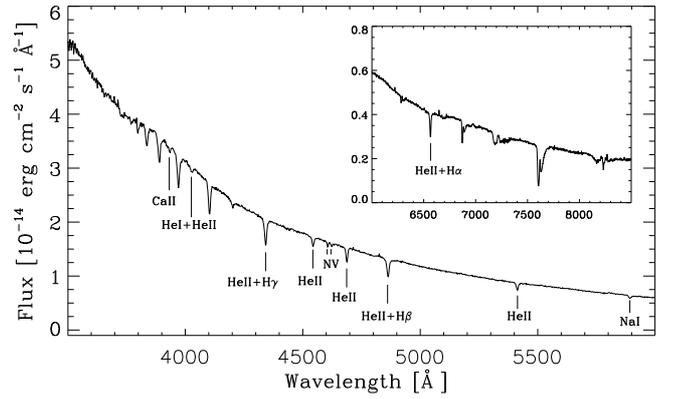}
      \caption{CAFOS CAHA blue spectrum of 2M1931+4324 in the range 3500--6000
      $\AA$. The inset shows the CAFOS CAHA red spectrum. Some helium, 
      hydrogen, and N\,{\sc v} absorption lines are indicated as well as the 
Ca\,{\sc ii}\,$\lambda$3968 and Na\,{\sc i}\,$\lambda$5895 in absorption.}
   \end{figure}
%

   \begin{figure}
   \centering
   \includegraphics[width=8.5cm]{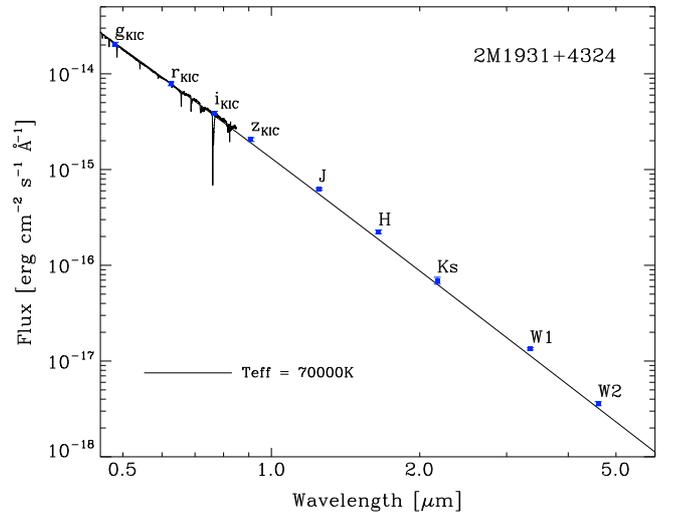}
      \caption{Spectral Energy Distribution (SED) of 2M1931+4324 constructed with the magnitudes from the Kepler Input Catalog (KIC), 
2MASS Point Source Catalog, and WISE archive. The blue and red CAFOS spectra (see Fig.\,4) are also included. The solid line 
corresponds to a blackbody with {\it T}$_{\rm eff}$ = 70000\,K, reddened by {\it A}$_{V}$ = 0.04 (see text) and normalized to the 
$r_{\mathrm{KIC}}$ magnitude.}
   \end{figure}
%

\subsubsection{The spectrum of 2M1931+4324}

Figure\,4 shows the optical spectrum of 2M1931+4324. It is dominated by strong ionized helium 
and hydrogen Balmer absorptions that are blended with the corresponding He\,{\sc ii} 
Pickering lines. The narrowness of the absorption lines and the presence of the He\,{\sc ii}$\lambda$4686 
line indicate a sdO nature for 2M1931+4324, in agreement with the classification by $\O$stensen et
al. (2010). We also note the presence of N\,{\sc v}$\lambda$$\lambda$4604,4620 absorption lines  
that are observed in some sdOs (Husfeld et al. 1989; Rauch et al. 1991). 
The lack of He\,{\sc i}$\lambda$4471 in absorption in the spectrum 
suggests a {\it T}$_{\rm eff}$ $\geq$ 60000\,K (Rauch et al. 1991) and,
therefore, that 2M1931+4324 is the central star of the nebula. Finally, a
careful inspection of the spectrum reveals Ca\,{\sc ii}\,$\lambda$$\lambda$3934 and 
Na\,{\sc i}\,$\lambda$$\lambda$5890,5895 absorption lines. These absorptions lines could be attributed to 
a late type companion, although they also are strong features of the interstellar medium. Nevertheless, the reddening 
towards the object is very low (see Sect. 2.3.2), which seems to favor the first possibility. A radial velocity analysis based 
on high-resolution spectroscopy of the central star will certainly allow us to decide between these two possibilities.

To obtain more information about 2M1931+4324, we have constructed its Spectral Energy Distribution (SED) 
with the available photometry: the g, r, i, and z magnitudes from the Kepler Input Catalog (KIC), the 
$J$, $H$, and $K_s$ magnitudes from the 2MASS Point Source Catalog\footnote{\url{http://irsa.ipac.caltech.edu}}, 
and the W1 and W2 magnitudes from the WISE archive. Figure\,5 presents the SED, including our optical spectrum. 
Because the red part of the SED is barely sensitive to {\it T}$_{\rm eff}$ for hot central stars, we have plotted 
on Fig.\,5 a blackbody with {\it T}$_{\rm eff}$ = 70000\,K, as a guide of what could be expected from extrapolating the stellar continuum of 
our red spectrum. The blackbody has been reddened by {\it A}$_{V}$ = 0.04 (corresponding to c(H$\beta$) = 0.02, see Sect. 2.3.2) 
and normalized at the $r_{\mathrm{KIC}}$ magnitude. Fig.\,5 shows that the KIC, near- and mid-infrared magnitudes agree reasonably 
well with that expected from a blackbody with {\it T}$_{\rm eff}$ $\geq$ 60000\,K. Moreover, no noticeable infrared 
excess can be recognized in the SED of 2M1931+4324. This result seems to rule out a (late type) subgiant or giant as 
the companion, pointing out to other possibilities (e.g., main sequence star, substellar companion). These possibilities 
should be compatible with the presence of Ca\,{\sc ii} and Na\,{\sc i} absorption lines 
in the stellar spectrum, if these lines have an stellar origin. A more detailed analysis of the SED will be possible 
once the atmospheric parameters of 2M1931+4324 have been accurately determined.

\subsubsection{The nebular spectrum}

In the nebular spectrum obtained at s1 (Fig.\,1), only faint H$\alpha$, H$\beta$ and [O\,{\sc
  iii}]$\lambda$$\lambda$4959,5007 emission lines are detected. Figure\,6 presents the
integrated spectra around these lines. The underreddened line intensities and their poissonian
errors are listed in Table\,1. They have been obtained using the extinction law
of Seaton (1979) and a logarithmic extinction coefficient
$c$(H$\beta$) $\simeq$ 0.02 derived from the observed H$\alpha$/H$\beta$ ratio,
assuming Case B recombination ($T_{\rm e}$=10$^{4}$\,K, $N_{\rm e}$=10$^{4}$\,cm$^{-3}$) 
and a theoretical H$\alpha$/H$\beta$ ratio of 2.85 (Brocklehurst 1971). 

The [O\,{\sc iii}]/H$\beta$ line intensity ratio of $\simeq$1.6 (Table\,1) indicates a 
very low-excitation PN. However, the lack of prominent low-excitation emission lines, in particular 
due to [N\,{\sc ii}], is highly peculiar. In fact, (low-excitation) bipolar
PNe usually present strong [N\,{\sc ii}] line emission 
that is not detected from the bipolar shell of PN\,G\,075.9+11.6. It could be that 
PN\,G\,075.9+11.6 is a density bounded PN with no low-excitation region or a very high-excitation 
PN with very weak [O\,{\sc  iii}] line emission due to prevalence of O$^{3+}$
in the nebula. Although the existence 
of (very) high-excitation is suggested by the outer filament, the absence of other 
high-excitation emission lines (e.g., He\,{\sc ii}, [Ar\,{\sc  v}], [Ne\,{\sc
  iv}]) in the optical spectrum makes these possibilities
questionable. Alternatively, the nebula could be strongly deficient in heavy
elements. This possibility would be compatible with both low- and
high-excitation in the nebula, and seems to be the only one that
accounts for the observed spectrum. A strong deficiency of heavy elements
would point out to a low-mass progenitor (say, $\leq$ 1\,M$_{\sun}$; see,
e.g., V\'azquez et al. 2002) for 2M1931+4324, which would be compatible with the relatively high 
Galactic latitude of the object (see Stanghellini et al. 2002). A much deeper
nebular spectrum is needed to identify more
(presumably extremely faint) emission lines and to obtain elemental abundances. 
 
The spectrum of the outer filament obtained at s2 (see Fig.\,1) is also shown in 
Fig.\,6. Only very faint [O\,{\sc  iii}]$\lambda$$\lambda$4959,5007 line emission is detected (observed 
[O\,{\sc  iii}]$\lambda$5007 flux $\sim$ 1.7$\times$10$^{-15}$ erg\,cm$^{-2}$\,s$^{-1}$\,$\AA^{-1}$), 
confirming its very high-excitation. Similar outer filaments and knots are
observed in other PNe (e.g, NGC\,40, Balick et al. 1992). 
In the case of PN\,G\,075.9+11.6, our data do not allow us to establish
whether the outer structures are related 
to mass ejection from 2M1931+4324 or they are ambient gas ionized by the
central star. Nevertheless, we note that 
the optical and mid-infrared images do not show the existence of nebulosities (except for those 
detected in our images) in the environment of PN\,G\,075.9+11.6.

\begin{table}
\caption{Emission line intensities in PN\,G\,075.9+11.6.}            
\label{table:1}      
\centering           
\begin{tabular}{lrc}   
\hline\hline           
Line & $f(\lambda)$ & $I(\lambda)$ $I$(H$\beta$)=100) \\   
\hline

H$\beta$$\lambda$4861          & 0.000    & 100 $\pm$ 3 \\

[O\,{\sc iii}]$\lambda$4959    & $-0.023$ &  37 $\pm$ 3 \\

[O\,{\sc iii}]$\lambda$5007    & $-0.034$ & 113 $\pm$ 3\\

H$\alpha$$\lambda$6563         & $-0.323$ & 285 $\pm$ 4 \\

\hline    

$c$(H$\beta$) = 0.02   \\ 

log$F$$_{\rm H\beta}$(erg\,cm$^{-2}$\,s$^{-1}$\,$\AA^{-1}$) = $-$14.31    \\

\hline                                  
\end{tabular}
\end{table}

   \begin{figure}
   \centering
   \includegraphics[width=8.5cm]{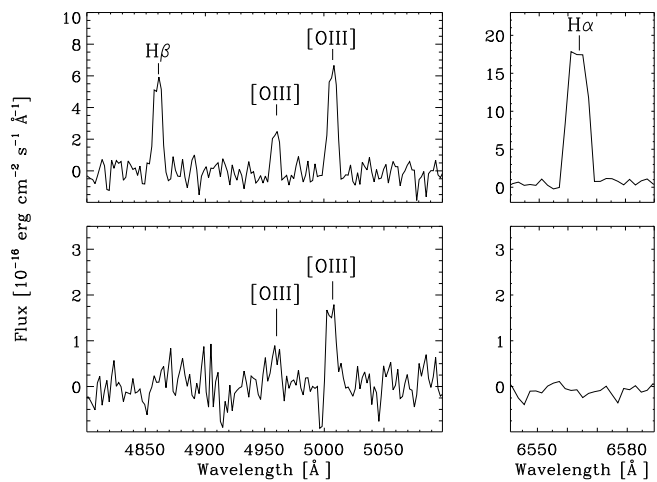}
      \caption{({\it top}) CAFOS CAHA spectra of the detected nebular emission lines from 
        PN\,G\,075.9+11.6. The spectrum has
        been obtained by integrating the detected emission lines at s1 (see
        Fig.\,1) between 38$''$ and 75$''$ north of 2M1931+4324. ({\it bottom}) Spectrum of the 
high-excitation filament. The spectrum has been obtained by integrating the detected 
[O\,{\sc iii}]$\lambda$5007 emission line at s2 (see Fig.\,1) in a region of 8$''$ centered on the filament.}
   \end{figure}

\subsection{High-resolution spectroscopy}

High-resolution long-slit spectra were obtained on 2012 May 14 and 15 with the Manchester Echelle 
Spectrometer (Meaburn et al. 2003) at the 2.1\,m telescope on the OAN 
San Pedro M\'artir Observatory\footnote{The Observatorio Astron\'omico
  Nacional at the Sierra de San Pedro M\'artir (OAN-SPM) is operated by the
  Instituto de Astronom\'{\i}a of the Universidad Nacional Aut\'onoma de 
M\'exico} (Baja California, Mexico). The detector was a 2k$\times$2k 
Marconi CCD that was employed with a 4$\times$4 binning, resulting in spectral 
and spatial scales of 0.11\,$\AA$\,pixel$^{-1}$ and 0$\farcs$702\,pixel$^{-1}$, respectively. 
A $\Delta$$\lambda$ = 90 {\AA } filter was used to isolate the 87$^{\rm th}$
order containing the H$\alpha$ emission line. The slit (6$'$ long, 2$''$ wide) was set on 
2M1931+4324 and spectra were obtained at two slit PAs: PA
58$^{\circ}$ (denoted S3) and PA 148$^{\circ}$ (S4). These slits are also shown in Fig.\,1. 
In the case of PA 58$^{\circ}$, two spectra were secured with the slit on
2M1931+4324 but displaced from each other 
along PA 58$^{\circ}$; these two spectra were combined in a single long-slit
spectrum. Exposure time was 1800\,s for each spectrum. A Th-Ar lamp was used
for wavelength calibration to an accuracy
of $\pm$ 1 km\,s$^{-1}$. The resulting spectral resolution (FWHM) is 12
km\,s$^{-1}$. Seeing was $\simeq$ 2$''$ during the observations. The spectra
were reduced with standard routines for long-slit
spectroscopy within the IRAF and MIDAS packages. 

   \begin{figure}
   \centering
   \includegraphics[width=8.5cm]{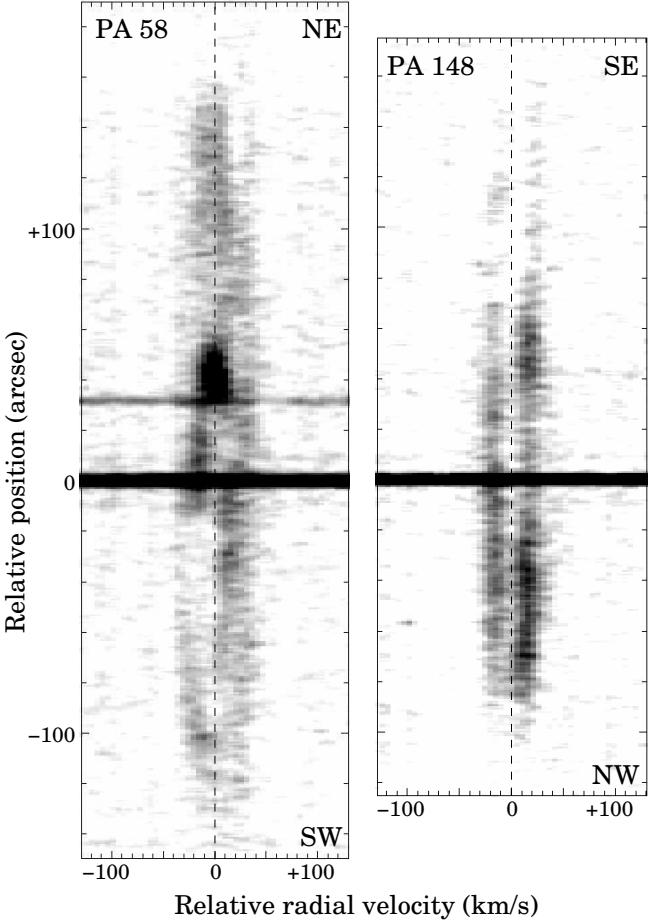}
      \caption{Grey-scale position-velocity maps of the H$\alpha$ line
        observed at two slit position angles (upper left). 
The origin (0,0) corresponds to the systemic velocity and the position of the
central star. The grey-scale is linear. A 3$\times$3 
box smooth was used for the representation.}
   \end{figure}
%

Figure\,7 shows position-velocity (PV) maps of the H$\alpha$ emission line at the
two observed PAs. Although the H$\alpha$ emission is weak, it shows details of
the internal kinematics of PN\,G\,075.9+11.6. From the velocity 
centroid of the line emission feature we derive a heliocentric systemic velocity 
of $-$16$\pm$2\,km\,s$^{-1}$. Internal radial velocities will be quoted hereafter 
with respect to the systemic velocity. 

The PV map at PA 148$^{\circ}$ (Fig.\,7) covers mainly the major axis of the elliptical shell.
It shows two extended velocity components with almost constant radial velocities of 
$\simeq$ $\pm$ 16\,km\,s$^{-1}$. The two components appear mainly parallel to the 
spatial axis in the PV map, although a slight tilt ($\simeq$
1--2\,km\,s$^{-1}$ in $\simeq$ 4$'$) could be present with the northwestern (southeastern)
region slightly blueshifted (redshifted). The PV map at PA 58$^{\circ}$ (Fig.\,7)
covers the major axis of the bipolar shell and shows a more complex kinematics. Two velocity 
components are observed at the stellar position with radial velocities of $\simeq$ $\pm$ 16\,km\,s$^{-1}$. 
Towards the southwest, the radial velocity of the two components increases up
to $\simeq$ $\pm$23\,km\,s$^{-1}$ at $\simeq$ 50$''$ from the 
central star and then it decreases until the two velocity components merge at
the systemic velocity, at the southwestern tip of the bipolar shell. An additional, faint velocity component
can be identified between $\simeq$ 20$''$ and $\simeq$ 80$''$ southwestern
from the central star, with radial velocities between $\simeq$ 10 and
$\simeq$ 0\,km\,s$^{-1}$, respectively. This component could be related to the
elliptical shell or indicate complex motions in the southwestern lobe of
bipolar shell. Towards the northeast, up to 35$''$ from the central star, two
velocity components are also recognized with radial velocities of $\simeq$
$-$15\,km\,s$^{-1}$ and $\simeq$ 25\,km\,s$^{-1}$. At $\simeq$ 40$''$ from 
the central star, a bright feature is observed at the systemic velocity, which
coincides with a bright region observed in the images (Fig.\,1) and may 
correspond to the northeastern edge of the elliptical shell. At angular
distances $>$ 50$''$ towards the northeast, the emission feature does not show
two velocity components but a single, relatively broad
feature mainly at the systemic velocity. This suggests that the northeastern 
lobe may contain material expanding a lower velocities than the
surface of the shell. Finally, the radial velocity of the northeastern tip of the 
bipolar shell coincides with the systemic velocity. 

The kinematics of the elliptical shell observed at PA 148$^{\circ}$ rules out that this shell
corresponds to the projection of a tilted ring, as the images could
suggest. If this was the case, one would not detect emission from the inner
regions of the shell, but only from the two ``points'' of the ring intersected by
the slit. The possibility of an oblate (equatorial) ellipsoid, as that
identified in Mz\,3 (Guerrero et al. 2004), may also be ruled out because one would 
expect to see in the PV map at PA 148$^{\circ}$ decreasing radial velocities
from the position of the central star to the tips of the elliptical shell,
which is not observed. The presence of two spatially extended velocity
components almost parallel to the spatial axis is compatible with a 
cylindrical (or open ellipsoidal shell) whose major axis is in (or close to) the
plane of the sky. The equatorial expansion velocity of this shell is $\simeq$
16\,km\,s$^{-1}$ and, if we assume homologous expansion 
(velocity proportional to radius), the expansion velocity at the observed
maximum size ($\simeq$ 2$\farcm$5, see Sect. 2.1) 
is $\simeq$ 45\,km\,s$^{-1}$. From the equatorial expansion velocity and the
equatorial radius of the elliptical shell ($\simeq$ 0$\farcm$9, Sect. 2.1), a 
kinematical age of $\sim$1.6$\times$10$^4$$\times$D[kpc]\,yr is obtained for this structure. 

The PV map at PA 58$^{\circ}$, along the main axis of the bipolar
shell, is compatible with the kinematics expected from bipolar motions in an
hour-glass-like shell.  The main axis of the bipolar shell should be located
in the plane of the sky, as indicated by the absence of tilt of the H$\alpha$
emission feature on the PV map. The equatorial expansion velocity of the bipolar shell 
is $\simeq$ 16\,km\,s$^{-1}$  while the polar one is $\simeq$ 43\,km\,s$^{-1}$
(also assuming  homologous expansion), which are virtually identical to those
of the cylindrical shell. From the equatorial expansion velocity and equatorial radius 
(0$\farcm$85, Sect. 2.1) of the bipolar shell, a kinematical age of
$\sim$1.5$\times$10$^4$$\times$D[kpc]\,yr is obtained for this structure,
which is very similar to that of the cylindrical shell. The kinematical ages
of the shells are compatible with an evolved PN. Moreover, the similarity
between the kinematical ages suggests that the formation of the two shells has
occurred in a short time span, as compared with the age of the nebula. 

\section{Discussion}

The detection of PN\,G\,075.9+11.6 around the sdO/central star 2M1931+4324
adds a new object to the known sample of sdO+PN associations and, therefore, a
new sdO whose origin can be ascribed to post-AGB evolution. It is worth noting that 
2M1931+4324 is quite bright ($r_{\mathrm{KIC}}$ $\simeq$ 13$\rlap.^m$9) as compared to most central stars of PNe. 
The fact that such a ``bright'' central star has gone unnoticed to date can be understood 
by its association with an extremely faint PN that cannot be recognized in the POSS with a 
simple visual inspection of the plates. In this respect, it should be mentioned that 
Jacoby et al. (2010) have recently identified a relatively large number of new PNe 
using the POSS plates, most of them at the detection limit of the plates. It is also noteworthy 
that PN\,G\,075.9+11.6 and a large fraction of the new PNe
identified by Jacoby et al. (2010) are located at relatively high Galactic latitudes 
($|{\it b}|$ $>$ 6$^{\circ}$) and, therefore, cannot be identified in recent PN
surveys that are concentrated more towards the Galactic plane (e.g., Parker et
al. 2005, Drew et al. 2005). A survey for PNe at higher Galactic latitudes 
may result in the identification of many new PNe (see Miszalski et al. 2011
and references therein) 

The existence of two shells in PN\,G\,075.9+11.6 indicates
that complex ejection processes have been involved in the formation of the
nebula. As already mentioned, the ejection of the two shells should have occurred in a
relatively short time span (as compared with the age of the nebula). Moreover,
the location of the two major axes, mainly in the plane of the sky, and the difference of $\simeq$
90$^{\circ}$ in their orientation indicate that the major axes are (virtually) perpendicular
to each other. Therefore, the central star has been able to eject
two shells in a relatively short time span, including, in addition, a large
change of $\sim$ 90$^{\circ}$ in the orientation of the main ejection axis. These
results, involving episodic ejections and changes in the orientation of the main
ejection axis, are difficult to interpret within a single star scenario but
fit in the framework of binary central stars in which they may be explained as a result
of stellar interactions, mass transfer and precession of the collimating
agent. In fact, PNe with multiple structures at different orientations, as
quadrupolar or multipolar ones, are usually interpreted invoking a binary
central star (e.g., Manchado et al. 1996; Guerrero et al. 2012 and references
therein). In this respect, the structure of PN\,G\,075.9+11.6 and the binary nature
of 2M1931+4324 (Jacoby et al. 2012) provide support for a binary star
scenario in multi-shell PNe and reinforce the idea that binary stars are
an important ingredient in the formation of complex PNe (De Marco 2009 and references 
therein; Miszalski et al. 2009; Miszalski 2012; Boffin et al. 2012). 

Within the context of PNe with binary central stars, we note the off-center
position of 2M1931+4324 with respect to the nebular shells. This situation has
already been observed in other PNe
(e.g., Sahai et al. 1999; Miranda et al. 2001) and interpreted as a result of
a possible binary central star (Soker et al. 1998 and
references therein). The binary nature of 2M1931+4324 provides support
for this interpretation. Peculiar in PN\,G\,075.9+11.6 is the very large
difference between the orientation of the two shells. Other quadrupolar or multipolar PNe do
not usually show such large differences in the orientation of the shells
(Manchado et al. 1996), although they have been observed in a few cases (see
Guill\'en et al. 2012, and references therein). In an axisymmetric PN with a binary central
star, the orientation of the main nebular axis may be
expected to be perpendicular to the orbital plane of the binary. If so, the
orbital plane of 2M1931+4324 should have been almost parallel to the line of sight (or
somewhat tilted, as eclipses are not observed, Jacoby et al. 2012) when each 
shell was formed, but it has rotated by $\sim$ 90$^{\circ}$ between the
ejections. This would imply a dramatic change in the angular momentum of the
binary star system, which is not easy to explain. On the other hand, the generation
of axisymmetric shells in PNe may be due to other mechanisms (collimated outflows,
magnetic fields, see, e.g., Balick \& Frank 2002) so that the main nebular
axis should not be necessarily related to the orbital plane of a binary central
star. Nevertheless, very large changes in orientation of the collimating agent
would still be required. In any case, PN\,G\,075.9+11.6 and its binary central
star 2M1931+4324 present characteristics that make this system another interesting
case to study the formation of PNe with binary central stars, as, e.g., ETHOS\,1 and Fleming\,1
(Miszalski et al. 2011; Boffin et al. 2012).

The role of the binary 2M1931+4324 in the formation of PN\,G\,075.9+11.6 could
be even more crucial if the nebula is strongly deficient in heavy elements and 
the central star evolves from a low-mass progenitor, as single stars with low
initial mass are expected to form spherical PNe but not axysimmetric and
multi-shell ones (e.g., V\'azquez et al. 1999, 2002; Stanghellini et al. 2002
and references therein). Besides obtaining elemental abundances in the nebula,
in order to confirm a possible deficiency of heavy elements, estimating the
atmospheric parameters of 2M1931+4324 will allow us to constrain the
initial mass of the progenitor.

\section{Conclusions}

Using deep H$\alpha$ and [O\,{\sc iii}] images we have detected a
very faint nebula around the sdO 2M1931+4324, recently found to be a 
binary star. The nebula presents a bipolar and an elliptical shell, as well as
high-excitation
structures outside the two shells. Faint emission from the central nebular
regions is also detected at 12\,$\mu$m and 22\,$\mu$m in archive WISE
images. Analysis of the internal nebular kinematics, by means of high-resolution, long-slit spectroscopy, 
reveals a bipolar 
shell and a cylindrical (or open ellipsoidal) shell with their major axes mainly 
perpendicular to each other. In addition, very similar expansion velocities are found in the
two shells which have been formed within a relatively small time span. 

Our intermediate resolution spectrum of 2M1931+4324 confirms its sdO
classification and indicates a $T_{\rm eff}$ $\geq$ 60000\,K, strongly
suggesting a PN nature for the detected nebula that is tentatively referred
as to PN\,G\,075.9+11.6. Therefore, 2M1931+4324 adds a new object to the known
sdOs associated with a PN and to the sdOs with a post-AGB origin. 

The spectrum of PN\,G\,075.9+11.6 exhibits only H$\alpha$, H$\beta$ and [O\,{\sc
  iii}] emission lines that indicate a very low-excitation ([O\,{\sc
  iii}] to H$\beta$ intensity ratio $\simeq$1.6), in strong contrast with
the absence of low-excitation emission lines. The possibility of a very
high nebular excitation is difficult to reconcile with the absence of other
high-excitation emission lines, suggesting that PN\,G\,075.9+11.6
might be deficient in heavy elements, a fact that should be confirmed
by means of very deep spectroscopy and elemental abundance calculations. 

The analysis of the spatiokinematical structure of PN\,G\,075.9+11.6 indicates
that 2M1931+4324 has been able to eject two axisymmetric shells in a
relatively short time span, as compared with the age of the
nebula of $\sim$1.6$\times$10$^4$$\times$D[kpc]\,yr, as derived in the present 
work. Between these two events, the main ejection axis has rotated by $\sim$
90$^{\circ}$ in such a way that the two main nebular axes are perpendicular to
each other and, in both cases, oriented almost in the plane of the sky. The
complexity of PN\,G\,075.9+11.6 and the binary nature of 2M1931+4324 provide
strong support to the idea that binary central stars are a key ingredient to
generate complex PNe.

\begin{acknowledgements}
We thank our anonymous referee for his/her comments that have been most helpful to improve the analysis and discussion
of the data. We are very grateful to G.\,Jacoby for sharing with us the result of the binary nature of 
2M1931+4324 and for useful comments. We also thank M. Tapia for fruitful comments. This paper has been
supported partially by grants AYA\,2009-08481, 
AYA\,2009-14648-02, AYA\,2011-30147-C03-01 and AYA\,2011-30228-C03-01 of the Spanish MINECO, and by grants
INCITE09\,E1R312096ES, INCITE09\,312191PR and IN845B-2010/061 of Xunta de Galicia, all of them
partially funded by FEDER funds. CR-L has a post-doctoral contract of the JAE-Doc program 
``Junta para la ampliaci\'on de estudios'' (CSIC) co-funded by FSE and acknowledges financial 
support by grant AYA\,2010-14840 of the Spanish MINECO. RV, LO, and PFG are supported by PAPIIT-DGAPA-UNAM 
grant IN109509. LO acknowledges support by project PROMEP/103.5/12/3590. Authors also acknowledge the  
staff at OAN--San Pedro M\'artir (particularly to Mr. Gustavo Melgoza-Kennedy), Calar Alto, and La Palma 
Observatories for support during observations. This research has made use of the SIMBAD database, operated at the 
CDS, Strasbourg (France), Aladin, NASA's
Astrophysics Data System Bibliographic Services and the Spanish Virtual
Observatory supported from the Spanish MEC through grant AYA2008-02156. This
publication makes use of data products from (1) the Wide-field Infrared Survey
Explorer, which is a joint project of the University of California, Los Angeles, and
the Jet Propulsion Laboratory/California Institute of Technology, funded by the
National Aeronautics and Space Administration, (2) from the 2MASS, which is a joint project of the 
University of Massachusetts and the Infrared Processing and Analysis Center/California Institute of 
Technology, funded by the National Aeronautics and Space Administration and the National Science 
Foundation, (3) from the Mikulski Archive for Space Telescopes (MAST). STScI is operated by the 
Association of Universities for Research in Astronomy, Inc., under NASA contract NAS5-26555. Support for 
MAST for non-HST data is provided by the NASA Office of Space Science via grant NNX09AF08G and other 
grants and contracts.
\end{acknowledgements}

\end{document}